\documentclass[sigconf]{acmart}
\usepackage{amsmath,amsfonts}
\usepackage{algorithmic}
\usepackage{graphicx}
\usepackage{textcomp}
\usepackage{xcolor}
\usepackage[flushleft]{threeparttable}
\usepackage{textgreek}
\usepackage{multirow,booktabs,siunitx}
\usepackage{pifont}
\usepackage{hhline}
\usepackage{graphicx}
\usepackage{xcolor,colortbl}
\usepackage{pgfplots, pgfplotstable}
\usepackage{spreadtab}
\usepackage{hyperref}
\usepackage{algorithm}
\usepackage{algorithmic}
\usepackage{tcolorbox}
\usepackage{subcaption}

\usepackage{xcolor}
\definecolor{light-gray}{gray}{0.95}
\newcommand{\code}[1]{\colorbox{light-gray}{\texttt{#1}}}

\def\rot{\rotatebox}

\AtBeginDocument{%
  \providecommand\BibTeX{{%
    \normalfont B\kern-0.5em{\scshape i\kern-0.25em b}\kern-0.8em\TeX}}}

\copyrightyear{2022} 
\acmYear{2022} 
\setcopyright{acmcopyright}
\acmBooktitle{30th International Conference on Program Comprehension (ICPC '22), May 16--17, 2022, Virtual Event, Pittsburgh, USA}
\acmConference[ICPC '22]{30th International Conference on Program Comprehension}{May 16--17, 2022}{Virtual Event, Pittsburgh, USA}
\acmPrice{15.00}
\acmDOI{10.1145/3524610.3527920}
\acmISBN{978-1-4503-9298-3/22/05}

\begin{document}

\title{Backports: Change Types, Challenges and Strategies}


 \author{Debasish Chakroborti}
 \affiliation{%
   \institution{University of Saskatchewan}
   \streetaddress{176 Thorvaldson Bldg, 110 Science Place}
   \postcode{S7N 5C9}
   \country{Canada}}
 \email{debasish.chakroborti@usask.ca}
 \orcid{0000-0002-1597-8162}

 \author{Kevin A. Schneider}
 \affiliation{%
   \institution{University of Saskatchewan}
   \streetaddress{176 Thorvaldson Bldg, 110 Science Place}
   \postcode{S7N 5C9}
   \country{Canada}}
 \email{kevin.schneider@usask.ca}
 
  \author{Chanchal K. Roy}
 \affiliation{%
   \institution{University of Saskatchewan}
   \streetaddress{176 Thorvaldson Bldg, 110 Science Place}
   \postcode{S7N 5C9}
   \country{Canada}}
 \email{chanchal.roy@usask.ca}

\renewcommand{\shortauthors}{Chakroborti and Schneider, et al.}

\begin{abstract}
Source code repositories allow developers to manage multiple versions (or branches) of a software system. Pull-requests are used to modify a branch, and backporting is a regular activity used to port changes from a current development branch to other versions. In open-source software, backports are common and often need to be adapted by hand, which motivates us to explore backports and backporting challenges and strategies. In our exploration of 68,424 backports from 10 GitHub projects, we found that bug, test, document, and feature changes are commonly backported. We identified a number of backporting challenges, including that backports were inconsistently linked to their original pull-request (49\%), that backports had incompatible code (13\%), that backports failed to be accepted (10\%), and that there were backporting delays (16 days to create, 5 days to merge). We identified some general strategies for addressing backporting issues. We also noted that backporting strategies depend on the project type and that further investigation is needed to determine their suitability. Furthermore, we created the first-ever backports dataset that can be used by other researchers and practitioners for investigating backports and backporting.
\end{abstract}




\maketitle

\section{Introduction}
\label{sec_intro}


Pull-based development in version control systems is crucial for obtaining contributions from both core and external members. A popular pull-based development platform is GitHub, where pull-request (PR) creation and submission are major parts of the development process. To contribute to an upstream repository (also known as an original repository) with a pull-request, a contributor first creates a forked repository (i.e., a personal copy). A pull-request can also be submitted by creating a branch. A branch also acts as an independent copy of the software for development inside a repository. In both cases, pull-requests can be submitted to any branch of an upstream repository. After submission, merging is considered by the reviewers and core members \cite{pullrequest} \cite{aboutbranches}. 

\begin{figure}
\includegraphics[width=.90\columnwidth]{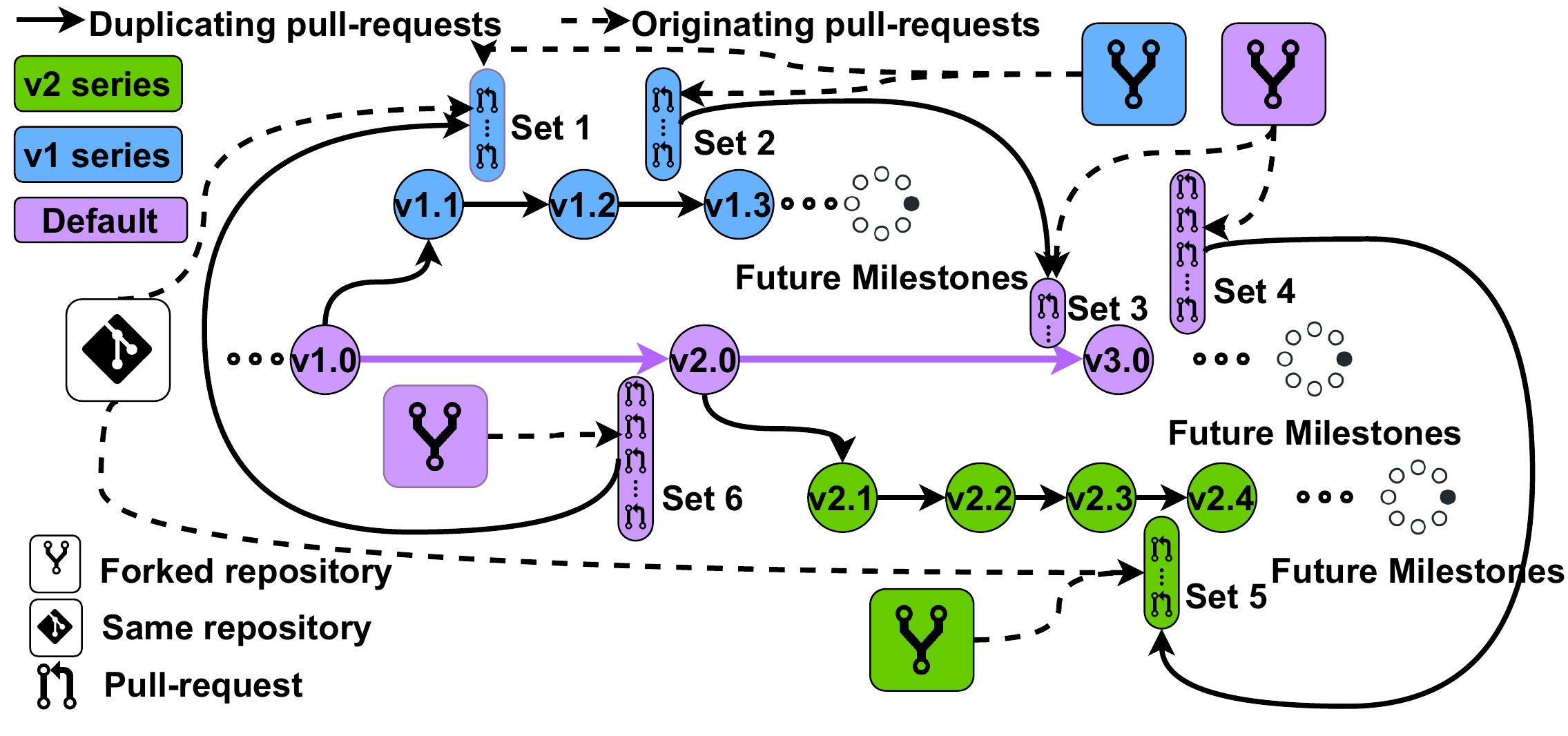}
\vspace{-3mm}
\caption{Backports and forwardports among different versions}
\label{fig:pull_network}
\vspace{-3mm}
\end{figure}

Merging software changes from one version to another is commonly known as porting. The term \emph{porting} includes changes from one software system to another, one repository to another, and one branch to another. Most commonly the term is used with \emph{commits}, and there is research \cite{7843626} \cite{10.1145/2393596.2393659}  \cite{6693095} \cite{10.1145/3338906.3342488} \cite{10.5555/3154690.3154693} on commit porting in cross-system management. We use the term \emph{backport} for merging changes from one repository to another or from the default branch to other branches for two reasons. First, in source-code repositories, a \texttt{default} branch always receives all the development activities and pull-requests, and \texttt{other} branches are just maintained (i.e., not actively developed) for different versions \cite{Kononenko_Water_2018_Shopify}. So, some changes from the current development branch (i.e., default branch) are always ported to previously defined version branches. Second, the term is commonly used among contributors in the open-source development community (i.e., used more than 162,189 times in 10 subject repositories). Thus, any changes from the \texttt{default} branch to \texttt{other} branches are called backporting changes in this paper. Similarly, any changes from \texttt{other} branches to the \texttt{default} branch are called forwardporting changes. Pull-requests related to these changes are commonly called backports and forwardports. Fig.~\ref{fig:pull_network} shows three development lines of the same software system: \texttt{default}, \texttt{v1}, and \texttt{v2}. Development of these different versions/branches can be pushed from different forked repositories. In this case, two forked repositories are created for the \texttt{default} branch, and two more are created for the \texttt{v1} and \texttt{v2} series. In the figure, dashed lines depict initialization of original pull-requests, and solid lines depict duplicating pull-requests to create backports/forwardports. Six sets of pull-requests are shown for three development lines. Each set represents all the pull-requests to a branch at a particular time and may contain both pull-requests and backports/forwardports. For example, Set 1 has three pull-requests, two are normal pull-requests, and one is a backport from the \texttt{default} series to the \texttt{v1} series after the \texttt{v1.1} version from Set 6. Similarly, Set 3 has forwardports from Set 2 and Set 5 has backports from Set 4 for the \texttt{default} branch and \texttt{v2} series, respectively. Thus, backports/forwardports can be generated for different versions; details are discussed in Section \ref{sec_idea}.


Previously, techniques were proposed for porting commits, specifically, for commit merging among cross-systems \cite{6693095} \cite{10.1145/2393596.2393659}, but it is hereto unknown how porting is done in a social coding environment(e.g., GitHub). Similarly, how porting is related to pull-requests is unexplored. 
This work explores a part of backport management with change types, strategies, and challenges in open-source software. In our exploration, we aim to answer the following research questions that can supplement porting knowledge to open-source software literature:

\textbf{RQ1}: \textit{How are pull-requests backported?}
Changes can be ported in many ways among versions. If they are ported with pull-requests they can be identified by the references/links. Both practitioners and researchers will be benefited from the knowledge of porting ways and intuition of versatile techniques to reference pull-requests.  

\textbf{RQ2}:  \textit{What kinds of pull-requests are backported?}
Tags are used in repositories for categorizing and organizing pull-requests. Backports are also tagged with various change types. It is important to know the common changes in backports before making decisions on them in repositories.  

\textbf{RQ3}:  \textit{What strategies are used for backports?}
Most of the repositories maintain contribution documents for instructing both core and community members. Investigating strategies with data can reveal current challenges from the required tasks to propose robust techniques for future management. 

\textbf{RQ4}:  \textit{What challenges are encountered when backporting?}
This research question will help us to improve the backporting process for practitioners. Moreover, future research directions can be identified for overcoming the challenges.   


To answer our research questions, we identified and analyzed 68,424 backports from 361,514 pull-requests in 10 subject repositories
and found that there are shortcomings in both research and management of backports. The main two stages of the backporting process are (1) identifying backports and (2) merging backports. There are some tools for identifying backports, but they are logic-based (e.g., pull-requests from one branch to another branch are considered to be backports) \cite{ansibledoc1} \cite{kibanadoc}. For merging, existing commit porting techniques \cite{6693095} \cite{10.1145/2393596.2393659} can be used, but how best to incorporate those techniques has not been determined. Since there are inconsistencies (49\%), incompatibles (13\%), delays (16 days to create, 5 days to merge), and failures (10\%) in backporting, our future work will be to overcome the problems in both identifying and merging backports. Towards identifying backports, we created a dataset for others to use. 

Our contributions can be summarized as follows: 

1. We identify backporting processes and change types.  

2. We explore and summarize the strategies of backporting.  

3. We identify challenges in strategies for our future goals. 

4. We created a dataset to investigate backports and for sharing with others.

The paper is organized as follows. Section \ref{sec_relatedwork} presents related work. Section \ref{sec_idea} discusses backports. Section \ref{sec_datacollection} presents our data collection process. Section \ref{sec_change_types} presents backport change types. Section \ref{efficiencies} discusses backporting strategies. Section \ref{sec_practice_challenges} discusses challenges in backporting. Section \ref{sec_result} answers the research questions. Finally, sections \ref{discussion} --\ref{sec_conclusion} conclude the paper with future directions and threats to validity.


\vspace{-3mm}
\section{RELATED WORK}
\label{sec_relatedwork}
Related studies explored recommending, analyzing, managing, porting, and classifying pull-requests.

\textbf{Recommending pull-requests.} 
A great number of works have been done on systems for recommending collaborators. Jiang et al. \cite{JIANG201748} analyzed the attributes of pull-based development and found that activeness is most important for commenter prediction. Similarly, Junior et al. \cite{JUNIOR2018181} analyzed attributes that can be suitable for integrator recommendations. Yu et al. \cite{YU2016204} \cite{7091328} analyzed attributes (i.e., comment network and traditional attributes) to propose a technique of reviewer recommendation.  Soares et al. \cite{SOARES201832} analyzed pull-request attributes to propose a reviewer recommender using association rules. Cheng et al. \cite{yang2018revrec} proposed an algorithm of recommender that recommends reviewers and reviewer roles. Junior et al. \cite{10.1145/2695664.2695884} proposed a technique of assigning developers when their number is low. Our work also analyzes attributes of pull-based development that are related to backports. Here, we explore a part of overall pull-requests (i.e., backports) to investigate change types, strategies and challenges.

\textbf{Analyzing pull-requests.} 
A number of studies analyzed pull-requests. Silva et al. \cite{10.5555/3021955.3021997} shows technical debt increases pull-request discussion. Rahman et al. \cite{10.1145/2597073.2597121} reveals common topics among successful and failed pull-requests. Kononenko et al. \cite{10.1145/3183519.3183542} \cite{Kononenko_Water_2018_Shopify} found size, people, experiences, and affiliation are important factors for merging decision and time. Terrell et al. \cite{terrell2017gender}  found that female contributions tend to be accepted more when gender is not specified. Yu et al. \cite{7180096} showed that commit size matters for reviewing. Zampetti et al. \cite{Zampetti_Penta_2019Interplay} showed that build-status has limited influence on pull-request closing. Gousios et al. \cite{icse16} found responsiveness, project compliance, communication are some major challenges of contributions from social, code, and tools aspects. Similarly, authors \cite{icse15} found that time, maintaining quality, reviewing, and rejection are common challenges for integrators. 
Similar to these studies, in our study, we also analyze pull-requests where code changes are intended for various versions. Contrary to the previous studies where challenges are investigated for practitioner perspectives, we identified backporting challenges from the metadata of their usage.

\textbf{Managing pull-requests.} 
Many studies investigated the organization of pull-requests. Li et al. \cite{10.1145/3131704.3131725}, Wang et al. \cite{10.1145/3361242.3361254}, Zhixing et al. \cite{Zhixing_Huaimin_2017_Duplicate} proposed techniques for detecting duplicate pull-requests using the title, description, and time factor. Zampetti et al. \cite{7961501} found that online resources are rarely referenced in pull-requests. Mohamed et al. \cite{8719563} predicted probable reopened pull-requests. Chen et al. \cite{10.1109/ICPC.2019.00037} predicted the acceptance of pull-requests using crowed source knowledge. Coelho et al. \cite{coelho2021empirical} distinguished refactoring-inducing from normal pull-requests in terms of no. of commits, time to merge and so on. Our study directs us to see some key parts of backporting that need to be managed to reduce inconsistencies, incompatibilities, delays, and failures.

\textbf{Classifying pull-requests and issues.} 
A number of studies have been done on classifying pull-requests in version control systems. Azeem et al. \cite{10.1145/3379177.3388904} classified pull-requests for integrators to accept, reject, and respond. Li et al. \cite{Li2017WhatAT} \cite{Li_Wang_2017_AutomaticRC} proposed a review-comment classification algorithm for understanding pull-requests. Yu et al. \cite{Yu_Li_2018_Classification} proposed a pull-request classification model for labelling. Veen et al. \cite{Veen_Zaidman_2015_Automatically_Prioritizing} prioritized important pull requests by considering static and dynamic information. Hoang et al. \cite{10.1109/ICSE-Companion.2019.00044} \cite{8896061} proposed a deep learning-based approach that can classify patches into stable and unstable classes. Jiang et al. \cite{jiang129recommending} proposed a neural network-based tag classifier for pull-requests. Cabot et al. \cite{7081875} classified issue labels based on string-based similarities. Izquierdo et al. \cite{7081860} developed a visualization tool to show label co-occurrences. Kallis et al. \cite{KALLIS2021102598} proposed a tool, Ticket Tagger, based on the fastText model. Mondal et al. \cite{amitscam_2019} showed that natural language can be used to classify code changes. A major difference between these works and ours is that we identify backports from pull-requests and classify backports based on available metadata of changes. 
(see Section \ref{sec_change_types}).

\textbf{Porting commits.}
Some studies worked on commit porting. Although these studies are not related to a contemporary social coding environment, we included these here to compare them with our future interests. Li et al. \cite{7843626} proposed a semantic slicing algorithm that can identify history commits for program correction while porting. Ray et al. \cite{10.1145/2393596.2393659} proposed a porting analysis tool, Repertoire, which finds ported edits from BSD cross-systems. Furthermore, Ray et al. \cite{6693095} proposed an algorithm to characterize semantic inconsistencies for ported code. On the other hand, Ren \cite{10.1145/3338906.3342488} recently proposed a prototype for porting patches automatically. However, their analysis is based on only 200 patches, and full implementation is still unavailable. Apart from these, Lawall et al. \cite{10.5555/3154690.3154693} proposed two tools, Prequel and GCC-reduce, that can help developers to collect porting information from the Linux kernel. Rodriguez et al. \cite{7371961} proposed a backporting strategy using the program transformation tool Coccinelle that only works with C code for Linux Drivers. Shariffdeen et al. \cite{10.1145/3460319.3464821} proposed an approach of partial transformation to backport patches for Linux kernels.  Thung et al.  \cite{7816469} proposed an approach to recommend changes while backporting Linux device drivers. Decan et al. \cite{9540328} investigate backporting practices for package distributions - Cargo, npm, Packagist and RubyGems and found dependable packages are vulnerable in some versions. The authors emphasized only on the security vulnerabilities, where we investigate all types of backports. Also, the work is limited to the package dependency network, where we explore all the change propagation in a repository from pull-requests. 

Prior studies mostly focus on managing and understanding pull-requests and commits in version control systems. Our focus is on backports and understanding current strategies/challenges with their change types in backporting.

\vspace{-3mm}
\section{Backports}
\label{sec_idea}

\begin{figure}
\includegraphics[width=.90\columnwidth]{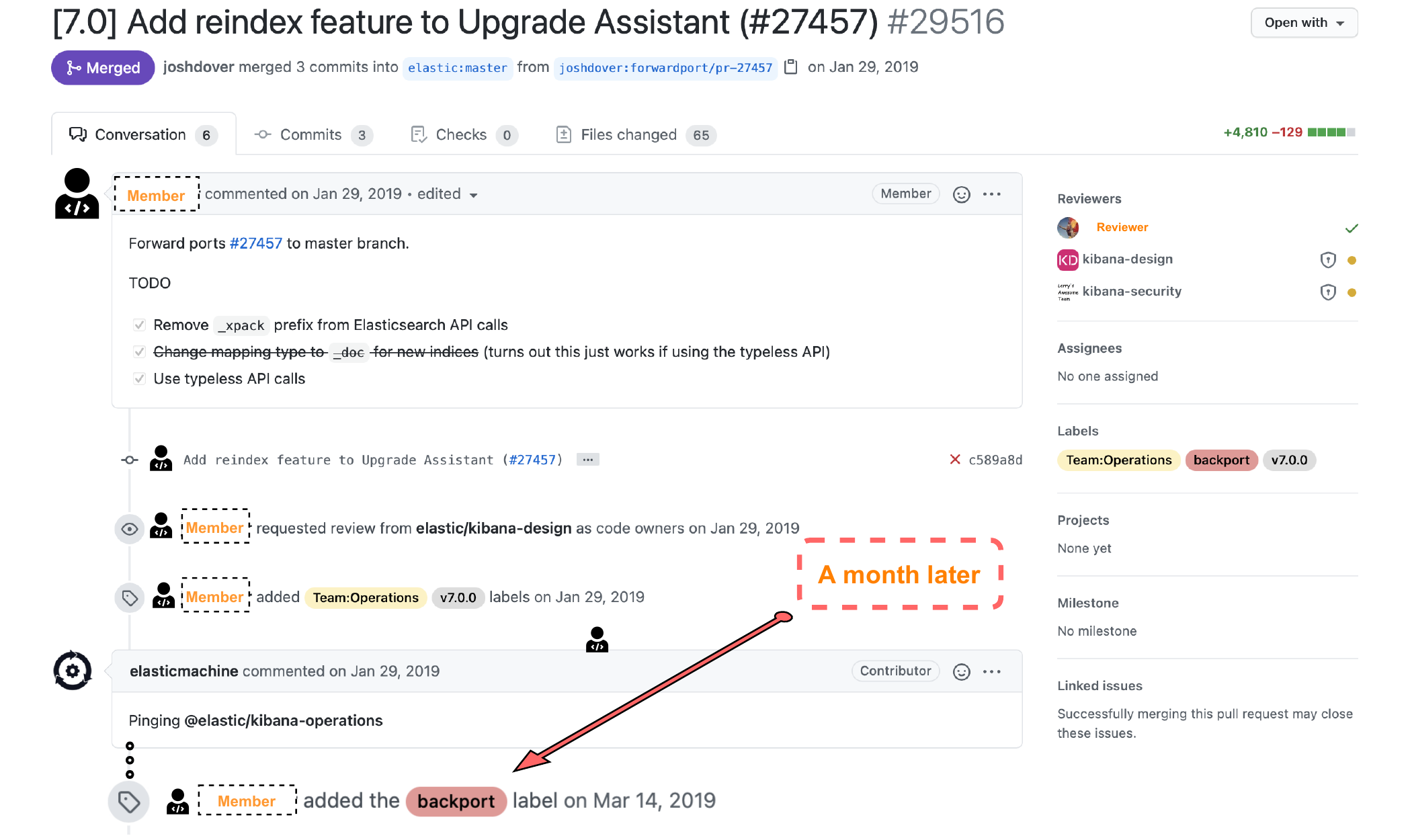}
\vspace{-2mm}
\caption{A pull-request for porting in Elasticsearch}
\label{exampleGithub}
\vspace{-3mm}
\end{figure}

This section provides backgrounds of backports for subject systems.  

\textbf{Backports to \texttt{other} branches.}  Generally, a change to the \texttt{default} branch (i.e., the main or current development branch) may be merged with \texttt{other} branches using a backport (i.e., we found 18.92\% of the total 361,514 pull-requests are backports). \texttt{Other} branches refer to any kind of independent line of software that originated from the \texttt{default} branch.
Our analysis suggests that in most cases, backports are requested by duplicating the original pull-requests from \texttt{other} branches of forked repositories to \texttt{other} branches of upstream repositories (i.e., 46,530 backports of 68,424, 68\% are requested from forked \texttt{other} branches to \texttt{other} branches) for propagating changes from main development line to versions. In addition, we found some backports where changes propagated between \texttt{other} branches in the same repository (i.e., 8\% of 68,424). The main reason to initiate backports from \texttt{other} branch to \texttt{other} branch by duplicating the work of original pull-requests is - the process is straightforward and merging needs less comparison. In this procedure, an immediate \texttt{other} branch (i.e., less deviated) is created to be a head, and the original \texttt{other} branch works as a base. Figure \ref{fig:pull_network} illustrates more forked repositories and pull-requests from \texttt{other} branches to \texttt{other} branches in six sets, where backports are initialized by duplicating original pull-requests.

\textbf{Forwardports to the \texttt{default} branch.} Our investigation identified a number of examples of merging backports to the \texttt{default} branch (i.e., 23\% of 68,424). For example, Figure \ref{exampleGithub} shows a forwardport to the master branch.  
Forwardports occur in two ways: 
(1) when a pull-request is merged with \texttt{other} branches, but the changes do not exist in the \texttt{default} branch; thus, later the pull-request is forward ported to the \texttt{default} branch; or, (2) when a pull-request has already been applied to the \texttt{default} branch, and after a discussion, someone tags it as a backport. In both cases, if a pull-request is merged to a \texttt{default} branch, then another pull-request can be created, or other merging methods can be used (e.g., a porting pull-request) for backporting the commit changes to the desired branch. We refer to original backporting related pull-requests in the \texttt{default} branch as forwardports (see Table~\ref{table_backport_forward_types}). Forwardporting is also referred to as reverse backporting, fore-porting and may even be considered an original pull-request.

\begin{figure}
\includegraphics[width=.85\columnwidth]{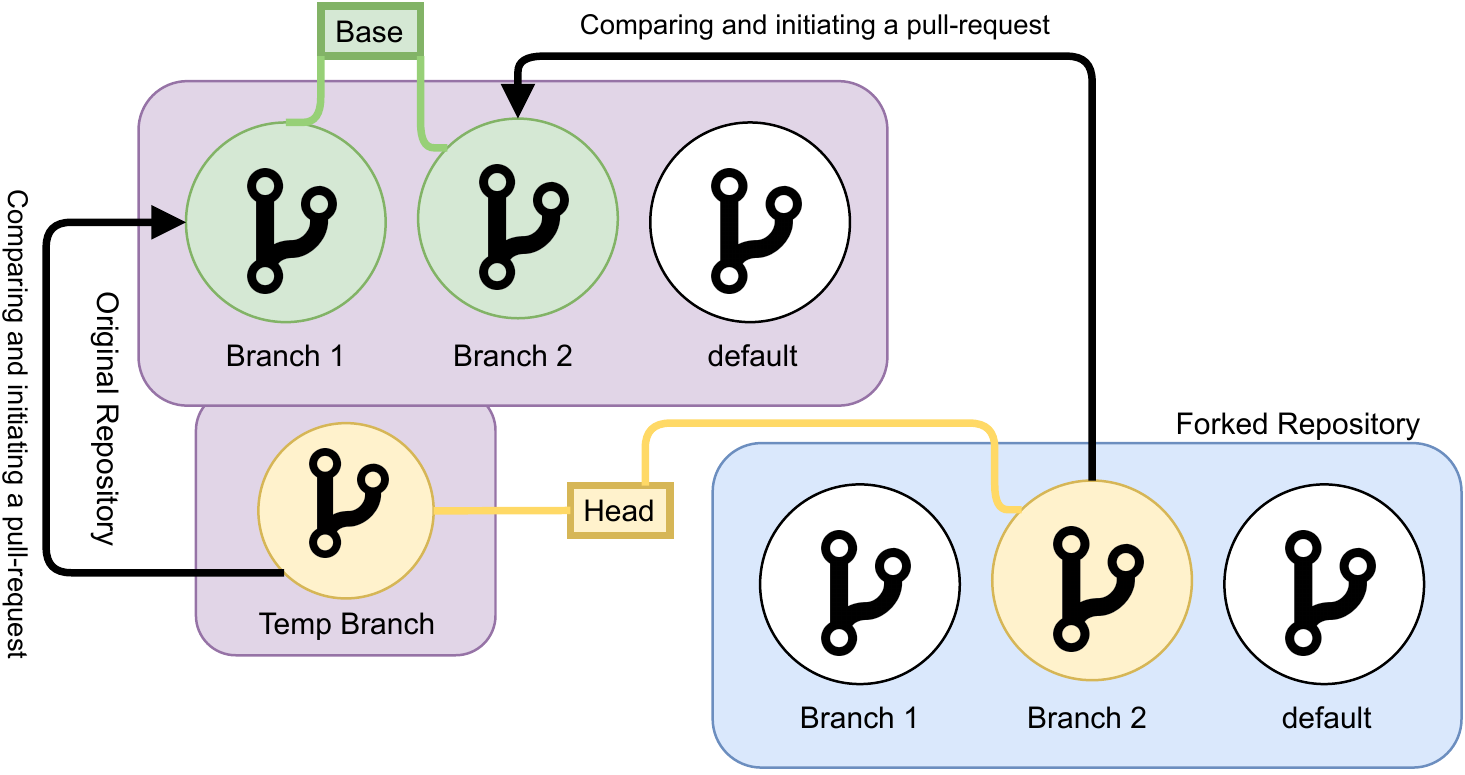}
\vspace{-3mm}
\caption{Head and base branches in repositories}
\label{headbase}
\vspace{-3mm}
\end{figure}

\textbf{Porting pull-requests.} A pull-request originates from a head branch to a base branch. Fig.~\ref{headbase} shows head and base information of both upstream and forked repositories. While comparing across existing branches or repositories, one can change the base to submit a pull-request to different branches. However, submitting a pull-request in this way can complicate the merging process as some of the requested changes will be unnecessary. Thus, creating a new branch (i.e., temp branch in Fig.~\ref{headbase}) or forking an updated upstream branch is easier for submitting and merging pull-requests. 

In addition to porting pull-requests, we also found that Git's cherry-picking (6.81\% of 68,424 backports talk about cherry-picking), rebase (1\% of 68,424), and patching commands (6.01\% of 68,424) can be used to merge backporting commits. Since these techniques are rarely used in our subject systems, we excluded them in findings for not being generalized for their low number.

\section{Data Collection}
\label{sec_datacollection}

\begin{figure*}
\pgfplotstableread{
row label backport normal
1 kubernetes 1108 61209
3 vscode 27 9036
4 nixpkgs 7659 89804
5 rust 2263 43000
7 origin 878 16253
8 ansible 8481 36140
10 corefx 178 24645
12 node 986 23211
13 tensorflow 57 16981
16 swift 92 36609
17 elasticsearch 20063 25058
18 moby 553 20550
19 cockroach 6603 24334
21 servo 29 16767
24 coreclr 100 17967
26 kibana 35416 28805
27 julia 1787 18264
28 TypeScript 24 12794
29 joomla-cms 171 20594
30 DefinitelyTyped 34 44994
34 amphtml 16 21074
35 edx-platform 70 26953
36 pandas 2110 18498
37 istio 447 18580
38 manageiq 6046 12245
39 godot 445 18296
40 gentoo 22 20120
43 magento2 2080 9853
44 salt 2337 33617
46 symfony 287 24849
47 cmssw 19714 10926
48 openshift-ansible 1345 8232
49 core 13568 6263
50 grpc 751 15551
53 gutenberg 257 15274
54 pytorch 33 34601
55 test-infra 41 18181
56 client 8 17522
57 react 36 10632
58 code-dot-org 237 39879
59 bitcoin 1067 13715
60 mbed-os 156 10205
61 scikit-learn 422 10229
62 server 9731 4166
63 charts 2 18125
64 terraform 532 10625
65 ant-design 1 9137
66 docs 0 2472
67 cms-sw.github.io 0 92
68 kafka 333 9423
69 electron 7881 5558
70 zephyr 1073 21477
71 wpt 5 25166
72 Marlin 7 9827
73 incubator-mxnet 395 9927
74 beam 130 14415
75 fastlane 2 6410
76 website 57 19856
77 rails 1747 25262
78 zulip 41 11439
79 framework 193 21140
81 gatsby 432 17024
82 crates.io-index 0 9
83 Guide_3DS 0 685
84 tidb 35 16831
85 angular-cli 28 7718
86 omim 2 12758
87 che 93 7147
88 browser-laptop 3 3661
90 components 11 10024
91 cpython 13717 10591
93 Cataclysm-DDA 24 30574
94 material-ui 76 12874
96 oppia 6 7410
97 alluxio 28 12162
98 XX-Net 0 478
99 cli 17 7380
102 react-native 32 9577
103 anime 0 125
104 create-react-app 6 3372
105 clipboard.js 0 177
106 nylas-mail 0 224
107 material-design-lite 4 1138
108 yarn 7 2081
109 swift 92 36609
110 angular.js 16 7881
111 Font-Awesome 1 587
112 animate.css 0 488
113 node 986 23211
114 go 0 1375
115 axios 0 727
116 electron 7881 5558
117 linux 0 440
118 d3 1 1101
119 javascript 2 1239
120 ohmyzsh 10 5474
121 bootstrap 438 11905
122 vue 8 1804
123 html5-boilerplate 0 1244
124 impress.js 0 320
}\mytable
\begin{tikzpicture}
\begin{axis}[
    ybar stacked,
    width = 1\textwidth,
    height=4cm,
  bar width=3pt,
    enlarge x limits=0.01,
    enlarge y limits=0.1,
    legend style={at={(0.45,0.85)},
      anchor=north, legend columns=-1},
    ylabel={\# of pull-requests},
    xlabel={Repository},
    xtick = {1,4,...,130},
    x tick label style={rotate=45, anchor=east},
    ]
\addplot+[ybar, black, draw=black, fill=black!20] table [y=normal, x expr=\coordindex] {\mytable};    
\addplot+[ybar, black, draw=black, fill=black!80] table [y=backport, x expr=\coordindex] {\mytable};
\legend{\strut Normal Pull-requests, \strut Backports}
\end{axis}
\end{tikzpicture}
\vspace{-3mm}
\caption{Number of backports and pull-requests for each repository}
\label{seriesRepo}
\end{figure*}
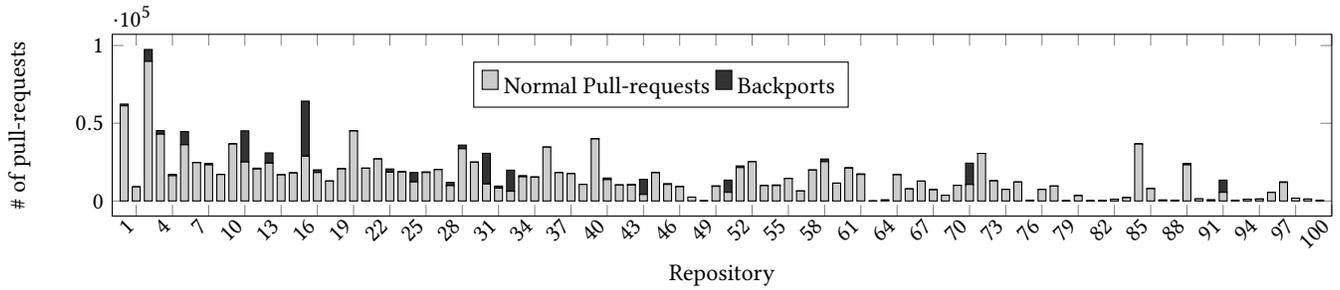

\begin{figure}
\includegraphics[width=.85\columnwidth]{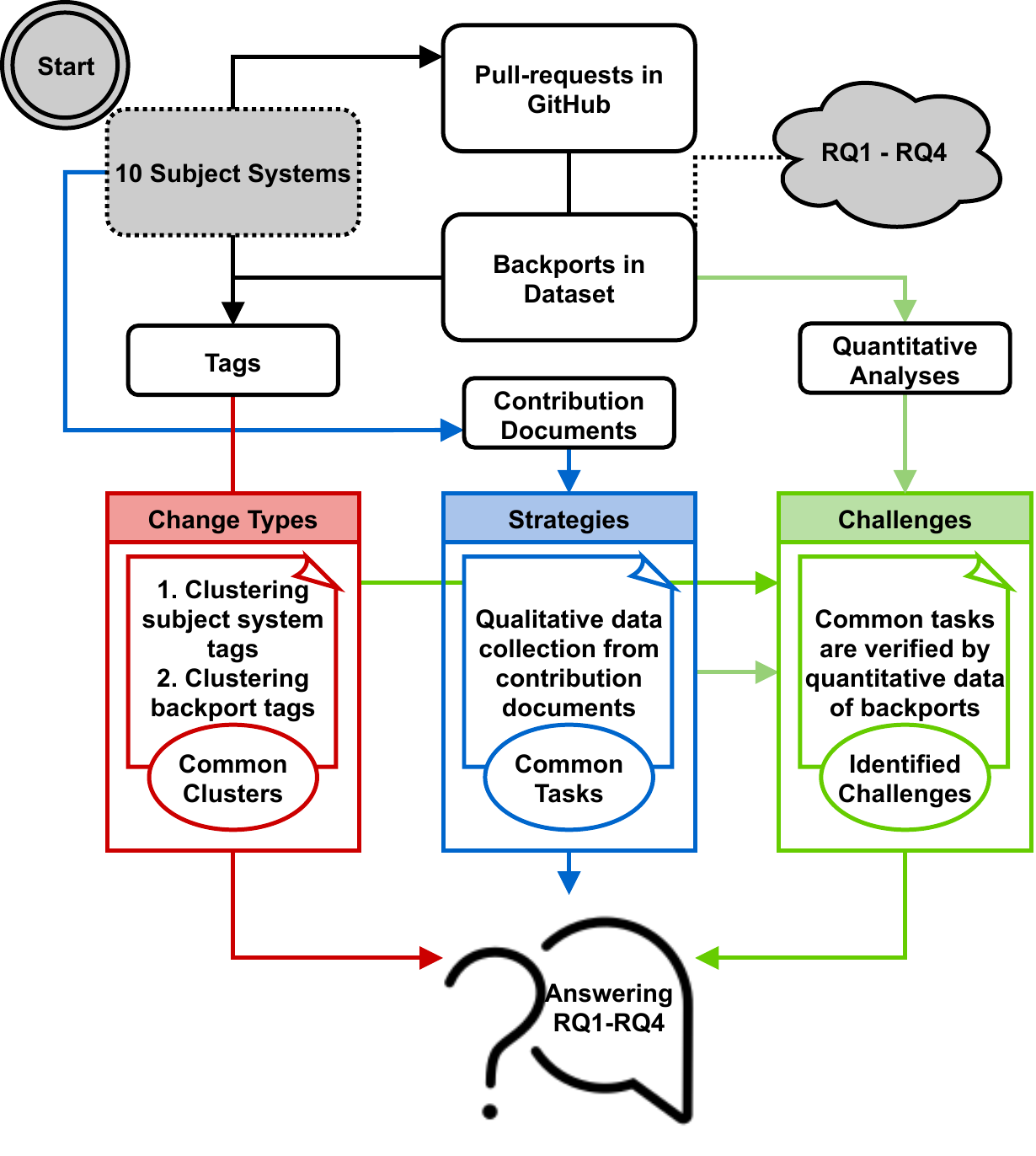}
\caption{Overview of the study}
\label{fig:methodology}
\end{figure}

We collected names of 1,000 top-ranked open-source projects from the Gitstar Ranking \cite{gitstar} website and randomly sampled 100 projects (since GitHub API calls are expensive) for our initial analysis. Figure~\ref{seriesRepo} shows the number of backports and pull-requests for these 100 projects.
 We match the keyword `backport' from the string representation of the pull-request's title, body, tags, and comments. If the keyword in label or title (i.e., already marked by human as a backport), we consider the pull-requests as backports. If the keywords are in body or comments, we consider having it two times (i.e., it is discussed two times for backporting) to identify pull-requests as backports. For our detailed analysis, we selected the top 10 projects (since calculating change percentages are expensive for commits, files, hunks, and lines) based on the number of backports (i.e., more than 400 to reduce the issues of marketing and advertising strategies shown by the work \cite{baltes2020sampling} and \cite{BORGES2018112}) that ensured a diverse range of projects and programming languages to support generalizing our findings. Default branches of the 10 repositories are identified from the repository settings and activities by the first author and validated by the second author while manually investigating the repositories. 
 From the 10 repositories, backports and forwardports are determined from their backport keywords and origin information (i.e., base and head), and we found 68,424 backports/forwardports. It should be noted that 79\% of the 68,424 pull-requests are confirmed backports from their tagged labels and the title prefix. For the others (i.e., not tagged with backport but has backport keywords), we selected 3 random sets with 10 pull-requests and found, on average, 73\% are backports. Thus, more than 90\% are confirmed backports in our dataset by the above heuristics. The GitHub API is used to collect the data as of April 15, 2021. An overview of the study is shown in Figure \ref{fig:methodology}.

\section{Backporting Change Types}
\label{sec_change_types}

Labels of pull-requests and issues are used to classify GitHub artifacts. There are two sets of labels in GitHub a default set of labels: \texttt{bug}, \texttt{duplicate}, \texttt{enhancement}, \texttt{help wanted}, \texttt{question}, \texttt{invalid}, \texttt{wontfix}, \texttt{good first issue}, \texttt{documentation} and custom set of labels: \texttt{repository-specific, created by team members}. Both sets are used similarly and are important for both issues, and pull-requests \cite{diniz2020github}. Pull-requests or issues are classified in many ways from their labels. Cabot et al. \cite{7081875} find \texttt{enhancement}, \texttt{bug}, \texttt{question}, \texttt{feature}, \texttt{documentation}, \texttt{wontfix}, and \texttt{task} are common labels for issues where they classify issue labels in \texttt{priority}, \texttt{version}, \texttt{workflow}, and \texttt{architecture} classes based on string similarities. Similarly, issue labels were presented with visualization of co-occurrences, user involvements and timelines by Izquierdo et al. \cite{7081860}. Jiang et al. \cite{jiang129recommending} classify pull-requests in five families - \texttt{Mark function}, \texttt{Give priority}, \texttt{Define status}, \texttt{Describe component}, and \texttt{Other} with a survey while proposing a tag recommender. There is work on labels that directly predicted labels rather than upper classes \cite{Abad2019LabelPO} \cite{KALLIS2021102598} \cite{diniz2020github}. Abad et al. \cite{Abad2019LabelPO} use text, description, and comment of issues to predict labels by support vector machine and naive Bayes multinomial. Similarly, Kallis et al. \cite{KALLIS2021102598} proposed the Ticket Tagger tool based on the fastText model that can tag issues as bug reports, feature requests, or questions. Jiang et al. \cite{jiang129recommending} also showed that tag usages are similar in issues and pull-requests. In addition, GitHub tags are used in issues and pull-request from the same corpus, where pull-requests are represented as issues for convenience. Here, we classify the labels of pull-requests/backports based on their semantic representation. We use Word Embedding in our classification algorithm since it can represent label meanings more accurately than graph-based models \cite{diniz2020github}.

\vspace{-3mm}
\begin{center}
\begin{table}[h]
\caption{Repository tags clustered by the algorithm}
\resizebox{.45\textwidth}{!}{%
\begin{tabular}{|m{1in}|m{1.2in}|m{1.2in}|}
\hline
\textbf{Top Cluster} & \textbf{Common keywords} & \textbf{Description} \\\hline
 Component, Team and Status (7) (10) (12) (13) & collection, aws, cisco, ansible, community, gcc, topic, merged, approve & mostly tags for modules \\ \hline
 Feature and Library (14) & feature, plugin, connection  & mostly tags for features \\ \hline
 Backport, Patch, Status (0) (9) & backport, inventory, patch, gem, approve, reject & mostly tags for porting and status \\ \hline
 Document and Component (6) (8)  & release\_note, doc, component, darwin, linux, cloudstack, css  & mostly tags for documenting \\ \hline
 Bug and Component (11) & bug, regression, break, fix, gcc & mostly tags for bug fixing \\ \hline
\end{tabular}
\label{table_repotypes}
}
\end{table}
\vspace{-3mm}
\end{center}

\begin{algorithm}
    \caption{Tag clustering algorithm}\label{cluster_tags}
    \begin{algorithmic}[1]
        \STATE  $\mathrm{tag\_Clustering} (tagFile)$
        \FOR{tags = $1$ to $COUNT(tagFile)$}
            \STATE tag\_sentences = Normalized tags at sentence level
        \ENDFOR
        \STATE model = $\mathrm{Word2Vec}(tag\_sentences)$
        \FOR{tags = $1$ to $tag\_sentences$}
            \STATE X\_tags =  $\mathrm{sentence\_Vectorizer}(tag\_sentence, model)$
        \ENDFOR
        \STATE clusters = $\mathrm{KMeansClusterer}(X\_tags)$
        \STATE Y\_tags = $\mathrm{TSNE}(X\_tags)$ 
        \STATE $\mathrm{Matplotlib}(X\_tags, Y\_tags, clusters)$
        
    \end{algorithmic}
\label{algorithm_cluster}  
\end{algorithm}
\vspace{-2mm}

\begin{figure*}
\centering
\begin{subfigure}{.45\linewidth}
    \centering
    \includegraphics[width=\textwidth]{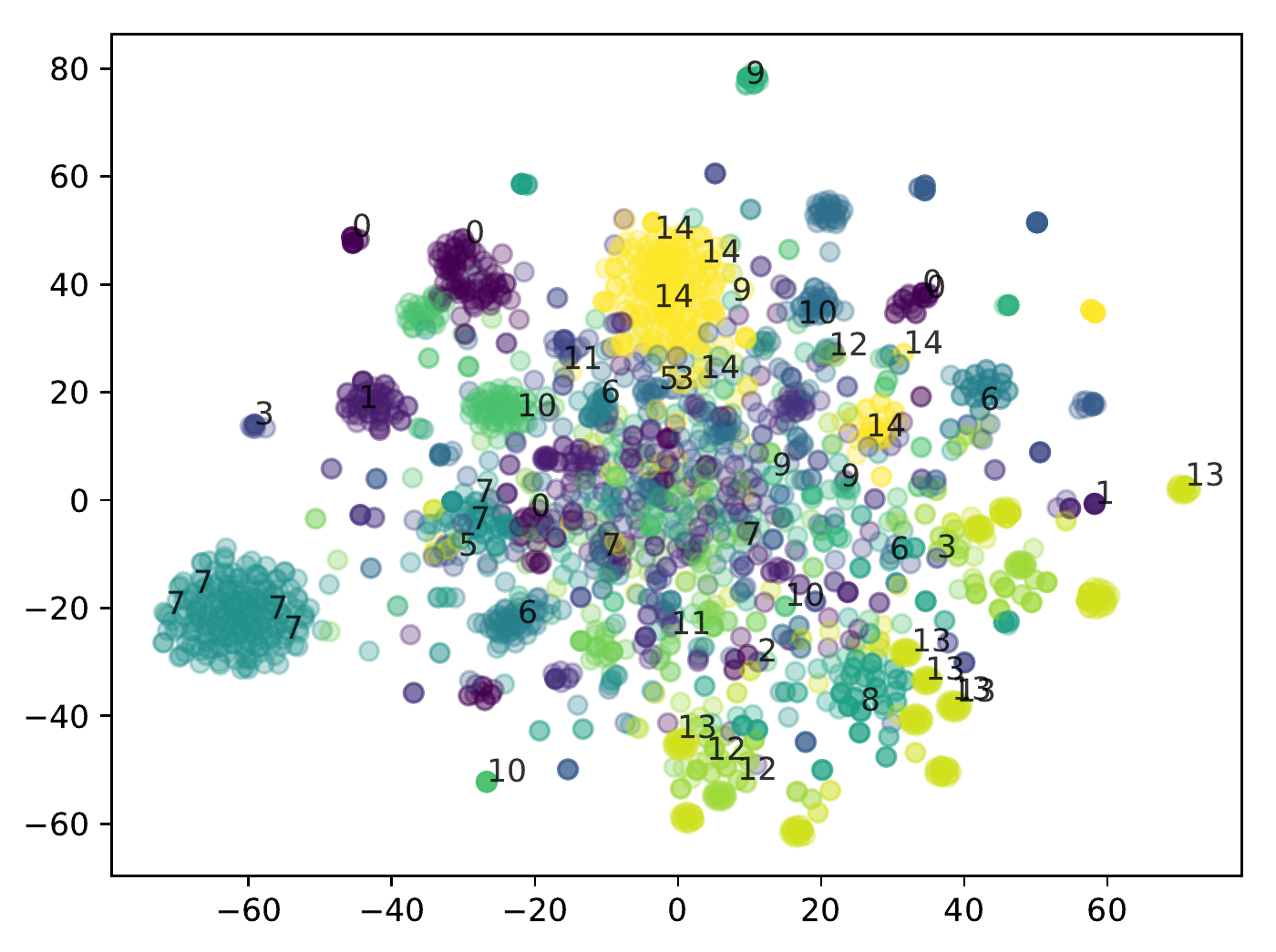}
    \caption{Repository tags}\label{fig:image11}
\end{subfigure}
\begin{subfigure}{.45\linewidth}
    \centering
    \includegraphics[width=\textwidth]{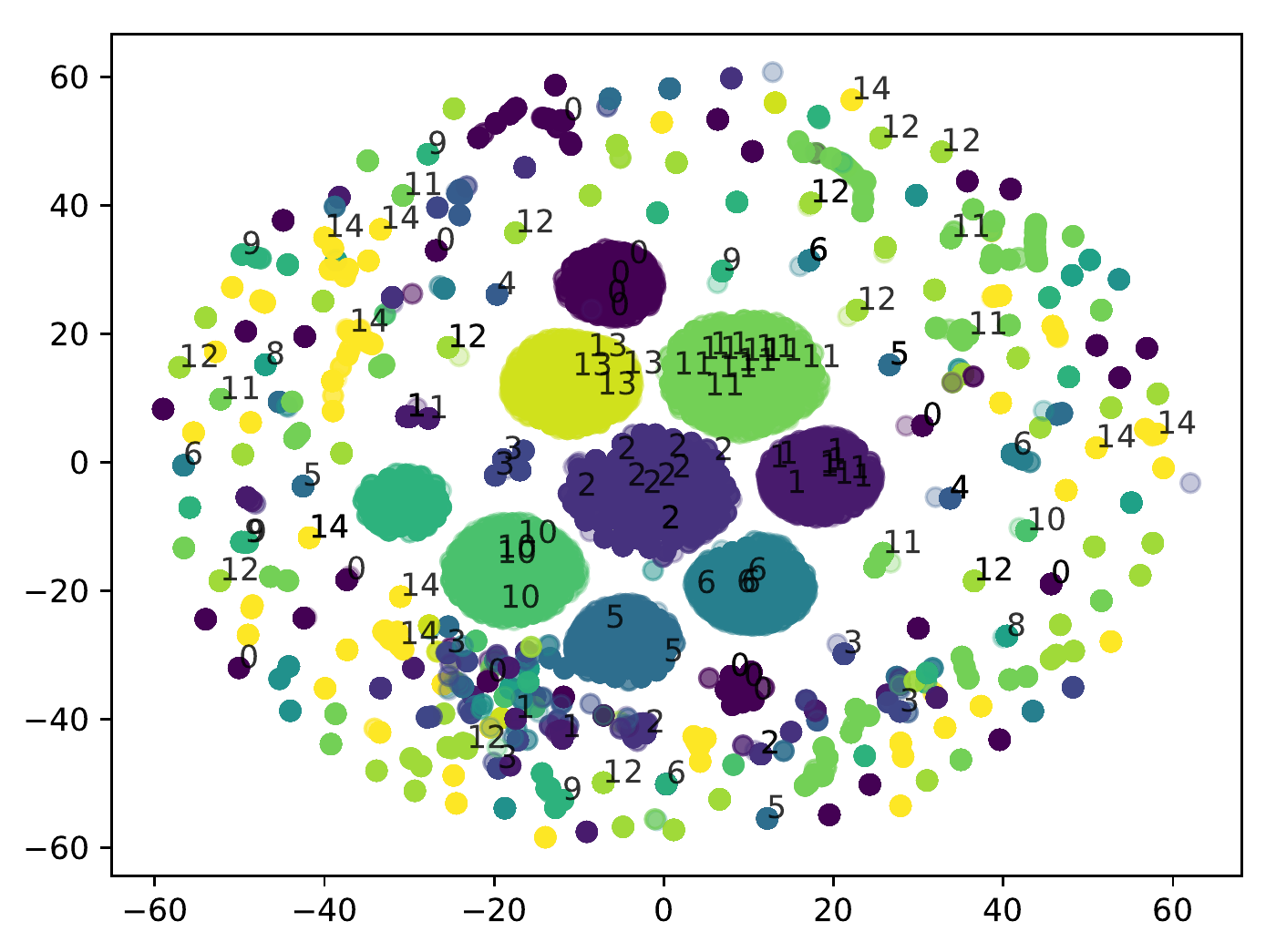}
    \caption{Backport tags}\label{fig:image12}
\end{subfigure}
\caption{Tag clustering of repositories and backports}
\label{fig:viz_clustering}
\end{figure*}

Algorithm \ref{algorithm_cluster} illustrates our tag clustering procedure. Since tags can have more than one word, we treated them as sentences and normalized them at sentence level in 1-4 lines. First, we tokenized the tags, removed punctuation, stop words, HTML tags/emojis, numbers, and at last lemmatized them. Normalized tags are fed in the Word2Vec model at line 5. Line 6 to 8 are used for averaging the tag word vectors at the sentence level. In line 9, KMeans algorithm \cite{hartigan1979algorithm} is used for tag clustering, where we kept K = 15 after trying K from 1 to 20. Line 10 and 11 are used for visualization. We used TSNE \cite{van2008visualizing} for reducing the data dimension and plotting as two dimensional. We included visualization in the algorithm since we assumed fully automated classification of software tags is not possible with natural language-based Word2Vec. From the visual representation, we identified top clusters with their assigned numbers. Later, we extracted all the tags in different files with respect to their assigned number. Then, we subdivided or aggregated a few clusters based on the software development meaning of the tags. Our algorithm is used in two stages - first, we gathered all tags of 10 subject repositories and clustered them with our algorithm. Figure \ref{fig:image11} shows clustering of all 3,094 tags of 10 repositories. In the figure, we can see that there are a few visible clusters, which are presented in Table \ref{table_repotypes}. In the table, tags are accumulated from 10 clusters (i.e., 0, 6, 7, 8, 9, 10, 11, 12, 13, 14). Here, we present 5 rows of the clusters after our manual intervention. For example, clusters (i.e., 7, 10, 12, 13) are accumulated together in the first row of Table \ref{table_repotypes} since the tags are overlapped semantically among them. We found many tags are used for identifying modules, components, development team, development stages (i.e., 1,601 of 3094 in Table \ref{table_repotypes}'s first row). For example, clusters (i.e., 7, 10, 12, 13) are accumulated together in the first row of Table \ref{table_repotypes} since the tags are overlapped semantically among them. In another cluster (i.e., 14 with 586 tags in the second row), we found tags are related to features and libraries. Other tags are mostly scattered with their semantic meaning. However, we found some backport, status and patch-related tags in clusters 0 and 9. Similarly, document and component tags are overlapped in clusters 6 and 8. Tags of bug and component are clustered in cluster 11. 
We identified a couple of clusters with common tags in repositories. We found that component tags are common in repositories with status, feature, library, document, backport, and bug tags.

\begin{center}
\begin{table}[h]
\caption{Backport tags clustered by the algorithm}
\resizebox{.45\textwidth}{!}{%
\begin{tabular}{|m{1in}|m{1.2in}|m{1.2in}|}
\hline
\textbf{Top Cluster} & \textbf{Common keywords} & \textbf{Description} \\\hline
 Bug, Security, Enhancement, Test and Documentation (1) (6) (8) (9) (11) (12) & bug, security, issue, enhancement, bug, fix, release\_note, authentication, analytic, doc, test, review, package, feature & mostly tags for changes \\ \hline
 Feature (14) & core, transform, feature, visualization & mostly tags for features \\ \hline 
 Version, Release and Refactor (2) & affect, refactor, revision, patch, sign & mostly tags for versioning and refactoring \\ \hline
 Component and Status (0) (3) & module, approve, networking, database & mostly tags for modules and statues \\ \hline
 Support (10) (13)  & core, community, support & mostly tags for core/remote teams \\ \hline
 Review and Integration (5) & core, review, ci, conflict & mostly tags for reviewing \\ \hline

\end{tabular}
\label{table_backporttypes}
}
\end{table}
\vspace{-3mm}
\end{center}

In the above clustering, we did not consider the usage of the tags; we considered their presence in repositories. The clusters showed us the common tag groups among the 10 subject systems. We found backport-related tags in the groups, and we explore them here as well as the common tags used with backports. 
Figure \ref{fig:image12} represents the clustering of the tags used with all the backports. Since we considered the usage of the tags, some commonly used tags are clustered in the middle of the figure. In Table \ref{table_backporttypes}, 13 clusters are accumulated in 6 rows. The first row represents 6 clusters (i.e., 1, 6, 8, 9, 11, 12), here we see different change types of pull-requests are clustered differently. The reason behind this is that repositories use different prefixes or postfixes in common tags. 
 In the first row's clusters, we found 30,912 out of 82,896 tags used with backports to represent bug, security, enhancement, test, and documentation type changes. The first three rows of Table \ref{table_backporttypes} show clusters for different changes.  The second row has around 4,332 tags with backports to indicate feature-related changes. The third row has around 6,969 tags for different versions and refactoring. As we found component and status tags are common in the repositories in the fourth row of Table \ref{table_repotypes}, we also see a significant number of tags are used with backports to indicate component and status in Table \ref{table_backporttypes}. Apart from these, support from remote/core teams, review, and integration scenarios are indicated with tags in backports (i.e., in fourth and fifth rows, respectively). Our target in Table \ref{table_backporttypes} is related to change types in backports, and they are in the first three rows. In this stage, again, we manually explored 8 clusters (i.e., 1, 2, 6, 8, 9, 11, 12, 14) to find common backport types. Here, backport types indicate the common changes in pull-request while porting them to other versions.  In eight clusters, we found eight common changes in backports, and they are fixing issues, securing code base, adding/improving test codes, adding/improving documentation, refactoring code, adding/improving features, optimizing performance, and others. Other type includes mainly tags related to modules, status, team, and releases as they are common in repositories. Similarly, backports are related to releases and versions, so tags of them are used while backporting. It should be noted that a pull-request/backport can have multiple tags and different changes. However, we found change-type tags are mostly used with module, release, and status tags in backports. Thus, the numbers in our results that might be overlapped are negligible with other change types. Here, we did not distinguish them and left as future work to identify the changes from the code level. Our findings of common changes in 8 clusters are illustrated in Table \ref{table_types}. In Table \ref{table_types}, we also gathered the keywords for similar changes to identify them in our experiments. 


\begin{center}
\begin{table}[h]
\caption{Change types identified by backport tags}
\resizebox{.45\textwidth}{!}{%
\begin{tabular}{|m{.45in}|m{1.4in}|m{1.4in}|}
\hline
\textbf{Type} & \textbf{Keywords} & \textbf{Description} \\\hline
 Fix & bug, break, fix, broken, error, fail, issue, fault, regression & issues   or   bugs    fixed \\ \hline
Security & encrypt, crypto, security, auth, ssl, tls, certificate, ssh & vulnerability patches \\  \hline
  Test & test, validation, verification & test script changes    \\ \hline
  Document & doc, note & improved documentation    \\ \hline
 Refactor & cleanup, refactor & structure improvements    \\ \hline
 Feature & feature, change, new, design,  visuali & new or improved functionality    \\ \hline
Optimize & enhance, performance, optimi& performance improvements    \\ \hline
Other & project specific keywords & usually used for team and module management    \\ \hline
\end{tabular}
\label{table_types}
}
\end{table}
\end{center}

\vspace{-3mm}
\section{Backporting Strategies}
\label{efficiencies}
In addition to identifying backport change types, we identified common backporting strategies. Our investigation of the repositories and available supporting documents (i.e., contribution documents collected from GitHub by the first author and discussed with the second author to find key elements) helps us identify a number of backporting strategies as written norms used by the subject repositories.
 Based on the strategies, we also analyzed the backports and the results are shown in Table \ref{table_backport_forward_types}, Table \ref{table_challengesbackports} and Table \ref{table_challengesBackportTypes}. 
 Table \ref{table_backport_forward_types} presents the results related to backports, forwardports, and backport change types. Table \ref{table_challengesbackports} presents the challenges of common tasks that we identified from the contribution documents. 
Challenges are also shown in Table \ref{table_challengesBackportTypes} for the backport change types we previously identified in Section \ref{sec_change_types}.  Details are discussed in Section \ref{sec_practice_challenges}. Here, we present excerpts from the contribution documents of our subject systems. 


\textbf{Ansible} \cite{ansibledoc1} \cite{ansibledoc2} uses a backporting script with \emph{pr\_id} and \emph{auto} parameters to indicate the original pull-request. Sometimes backport and core\_review labels are attached by a bot to each pull-request that is requested by \texttt{other} branches (i.e., include release branches). We found 91.92\% of 5,509 backports are tagged with backport label in the Ansible repository. Forwardporting pull-requests in the \texttt{default} branch first and creating changelog fragments (i.e., a YAML file) for bug-fix backports are encouraged. They discourage features from being backported to \texttt{other} branches. We also found 70.81\% backports of Ansible in Table \ref{table_backport_forward_types}  are related to fixing bugs. Although features are discouraged, around 8.77\% of backports are related to features. Table \ref{table_backport_forward_types} also shows backports related to documentation and testing are common in Ansible.  

Maintainers of \textbf{Rails} \cite{railsdoc} prefer to merge security and bug fixing commits in \texttt{other} branches by backports. Features are usually not backported unless there is confusion to separate them from bugs. Core team member resolves confusion after discussions. In a backporting process, commits are squashed for future maintenance. Merging of backports is done in two ways - small changes are backported by creating git patches, and separate pull-requests backport large changes. Similarly, maintainers of \textbf{OwnCloud} \cite{ownclouddoc}, and \textbf{Symfony} \cite{symfonydoc} prefer to create change-logs for backports since the backporting is highly related to releases. \textbf{Bitcoin} \cite{bitcoindoc} prefers for backporting bug and security fixing pull-requests in release branches, and backports are managed and merged in batches by core members. Figure \ref{fig:image12}'s cluster 2 and Table \ref{table_backporttypes}'s third-row show a large number of release tags are attached to backports.     

\textbf{ElasticSearch's} \cite{elasticsearchdoc} pull-requests are submitted to the \texttt{default} branch first; if merged, backports are created for the desired \texttt{other} branch. ElasticSearch row of Table \ref{table_backport_forward_types} show a significant number (i.e., 1,438) of forwardports that are merged in the \texttt{default} branch. Besides, Table \ref{table_challengesbackports} shows only 12\% inconsistencies for ElasticSearch, which is less than the average inconsistencies of 10 subject repositories. 

Maintainers of \textbf{Julia} \cite{juliaLangdoc}, \textbf{Nxpkgs} \cite{nixpkgsdoc}, and \textbf{Electron} \cite{electrondoc} repositories have several discussions to tag required pull-requests in \texttt{other} branches upon a release. Usually, tagged pull-request commits of Julia and Nxpkgs are cherry-picked to release branches. This might be the reason for the low number of backports in Julia and Nxpkgs repositories (i.e., 152 and 1,031) compared to their pull-request numbers in Table \ref{table_backport_forward_types}. On the other hand, 27.83\% of total pull-requests are backported in Electron with backports.  

In the \textbf{Kibana} \cite{kibanadoc} repository, backports are also merged into the \texttt{default} branch first. Maintainers also use a script (i.e., an automatic merging workflow) to merge labeled small backports automatically. In Table \ref{table_backport_forward_types}, we see a huge number (i.e., 26,888) of backports for Kibana repository. Besides, for automated maintenance their inconsistency (i.e., 10\%) and incompatibility (i.e., 03\%) is low in Table \ref{table_challengesbackports}.

In the \textbf{CMSSW} \cite{cmsswdoc} and \textbf{Cpython} \cite{cpythondoc} repositories, maintainers recommend submitting backports with original pull-request information and backport keywords in the title. As a result, both repositories have considerable numbers of backports compared to their total pull-requests, 2,540 and 1,038 in Table \ref{table_backport_forward_types}, respectively.

The preceding examples are for just a few repositories on GitHub. Nevertheless, backporting is a common mechanism used by most repositories to port changes among version branches. In Figure \ref{seriesRepo} we see that out of 1,500,810 pull-requests from 100 repositories, there are a total of 184,090 backports. Based on our analysis of the 10 subject systems' contribution documents, bug and security fixing backports are emphasized in most of the repositories. However, our analysis revealed other change types are frequently backported. The most common backport pull-requests apart from bug and security fixes are test, document, and adding/modifying small features. A few repositories discourage including new features in backports, but feature changes is one of the change types that is frequently backported. We need to investigate more why feature backports are discouraged but are backported in most of the repositories. Some repositories tried to automate parts of the backporting process to overcome some of the challenges identified in Table \ref{table_challengesbackports}. Another common scenario in the repositories we see is to first request backports to the \texttt{default} branch. 
Furthermore, backports are highly related to releases, and how they are affecting the milestones also needs to be investigated further. Overall, most of the repositories discuss and pay special attention to handling backports, where backports are initiated after the original pull-request merging.  


\begin{table*}
\centering
\caption{Appearance of backports, forwardports and change types}
\resizebox{.85\textwidth}{!}{%
    \begin{tabular}{|c|l|l|l|l|l|l|l|l|l|l|l|}           
    \hline 
  \multicolumn{1}{|c|}{\multirow{2}{*}{Repository with pull-request}}    &
  \multicolumn{1}{|c|}{\multirow{2}{*}{\# Porting (\# Normal)}}    &
  \multicolumn{2}{c|}{Porting Type}    &\multicolumn{8}{c|}{Change types in backports}  \\   \cline{3-12}
\multicolumn{1}{|c|}{} &\multicolumn{1}{|c|}{} &{Back} &{Forward} & {Fix} &{Security} &{Test} &{Document} & {Refactor} & {Feature} & {Optimize} & {Other}             \\   \hline 
\multicolumn{1}{|c|}{Ansible (44,048)} & 5,509 (38,539)   & 5,044   & 465   & 3,901   & 152   & 1,240   & 1,291   & 0  & 483 & 12 & 292 \\ \cline{3-12}
\multicolumn{1}{|c|}{Nixpkgs (97,100)} & 1,956 (95,144)     & 1,031   & 925     & 9      & 217  & 1   & 108   & 80 & 196 & 13 & 1,442 \\ \cline{3-12}
\multicolumn{1}{|c|}{Bootstrap (12,067)}& 725 (11,342)    & 72  & 653   & 7      & 0  & 6   & 215   & 0 & 33 & 2 & 469  \\ \cline{3-12}  
\multicolumn{1}{|c|}{ElasticSearch (43,757)}& 11,238 (32,519) & 9,800   & 1,438   & 1,769   & 256   & 822   & 1,840   & 223 & 1,381 & 551 & 5,556 \\ \cline{3-12}  
\multicolumn{1}{|c|}{Julia (19,628)}&  469 (19,159)   & 152   & 317     & 82     & 3  & 10  & 23  & 0 & 5 & 12 & 340 \\  \cline{3-12}  
\multicolumn{1}{|c|}{Cpython (24,412)}& 5,485 (18,927)      & 1,038   & 4,447   & 958      & 60   & 325   & 1,033   & 0 & 2,626 & 92 & 2,221 \\  \cline{3-12}  
\multicolumn{1}{|c|}{Electron (13,485)}& 3,881 (9,604)    & 3,753   & 128   & 0      & 19   & 0   & 20  & 0 & 162 & 1 & 3,679  \\  \cline{3-12}
\multicolumn{1}{|c|}{OwnCloud (19,775)}&  3,877 (15,898)    & 1,714   & 2,163     & 334      & 25   & 992   & 16  & 0 & 206 & 92 & 2,324 \\  \cline{3-12}  
\multicolumn{1}{|c|}{CMSSW (30,571)}& 4,602 (25,969)    & 2,540   & 2,062     & 146      & 0  & 4,586   & 3   & 0 & 68 & 0 & 16 \\  \cline{3-12}  
\multicolumn{1}{|c|}{Kibana (56,668)}&  30,682 (25,986)   & 26,888 & 3,794  & 475      & 299  & 169   & 3,487   & 4 & 1,789 & 231 & 26,453 \\  \cline{3-12}

 \hline 
 \multicolumn{1}{|c|}{Average}& 6,842.80 (29,308.70)   & \textbf{5,203.20} & \textbf{1,639.20} & 768.10 & 103.10 & 815.10 & 803.60 & 30.70 & 694.90 & 100.60 & 4,279.20 \\ \hline 
 \end{tabular}
 \label{table_backport_forward_types}
 }
 \end{table*}


\section{Backporting Challenges}
\label{sec_practice_challenges}

\begin{table}
\resizebox{.85\columnwidth}{!}{%
\begin{threeparttable}
\centering
\caption{Challenges in pull-request backporting}
\label{table_challengesbackports}
\begin{tabular}{|c|l|l|l|l|l|}
\hline 
 \rot{0}{Repository}    & \rot{0}{INC}      & \rot{0}{IPA}      & \rot{0}{Creating}      & \rot{0}{Merging}  & \rot{0}{Failport} \\
  \rot{0}{}    & \rot{0}{(\%)}      & \rot{0}{(\%)}      & \rot{0}{delay(days)}      & \rot{0}{delay(days)}  & \rot{0}{(\%)} \\ \hline
Ansible        & 53.11          & 13          & 13       & 06 (17)  & 10.44 (27.24) \\
Nixpkgs      & 60.63          & 18          & 21       & 02 (08) & 12.52 (15.00) \\
Bootstrap      & 65.51          & 25          & 22       & 15 (10) & 08.41 (46.77) \\
ElasticSearch  & 22.13          & 12          & 06        & 01 (05)  & 05.77 (18.21) \\
Julia          & 51.38          & 20          & 25       & 07 (12)  & 08.74 (16.91) \\
Cpython        & 96.80          & 16          & 30      & 10 (13) & 23.20 (12.97) \\
Electron       & 17.18          & 04          & 05       & 01 (05) & 05.43 (14.55) \\
OwnCloud       & 57.03          & 11          & 18       & 02 (07) & 12.04 (18.31) \\
CMSSW        & 56.93          & 12          & 17       & 10 (06) & 14.10 (18.12) \\
Kibana         & 10.24          & 03          & 01       & 01 (05) & 03.46 (15.31) \\ \hline
Average & 49.09 & 13.4 & 15.80 &  5.40 (8.83) & 10.41 (20.34)  \\
\hline
\end{tabular}
    \begin{tablenotes}
      \small
      \item Values in parentheses are for normal pull-requests.
      \item INC = Inconsistency, IPA = Incompatibility.
    \end{tablenotes} 
\end{threeparttable}
}
\vspace{-3mm}
\end{table}

Even though backporting is common among repositories, a number of backporting tasks need to be performed manually. We present these tasks with some challenges after investigating backport-related data. We are interested to see the link between backports and forwardports, the amount of changes required in backports, the time required to identify a backport, and failing information.

\textbf{Inconsistency.} Normally backports are created after a pull-request is merged to a \texttt{default} branch of an upstream repository. When creating a backport, it is typical to include a reference to the original pull-request. For example, Ansible backport \#76043 \cite{incon_pullrequest76043} references the original pull-request \#76041 \cite{incon_pullrequest76041}. However, we found that there are many cases when these references are missing (e.g., Ansible pull-request \#14565 \cite{incon_pullrequest14565}). We use the regular expression \verb!r'#[\d]+'! to extract a pull-request reference and verify the reference is an original pull-request with a GitHub API endpoint. In our analysis of the 10 subject repositories, we found that 49.09\% of backports (see Table~\ref{table_challengesbackports}) do not have this linking. In addition, change types in Table \ref{table_challengesBackportTypes} show inconsistencies between 33.22\% to 64.35\%. We can also get some ideas of the inconsistencies from Table \ref{table_backport_forward_types}. The third and fourth columns of Table \ref{table_backport_forward_types} show the number of backports and forwardports of each subject system. On average, we found 5,203.20 backports compared to only  1,639.20 forwardports. Table \ref{table_backport_forward_types} shows the number of backports for individual repositories is always higher than the forwardports except for Bootstrap, Julia, CPython, and OwnCloud Core where the \texttt{default} branch's pull-requests are highly maintained, or they do not have specific tags for backports. For example, OwnCloud does not have any explicit backport tag, and \texttt{default} branch pull-requests are discussed in comments for backporting. Bootstrap, Julia, and CPython use \emph{backport-to-vX} or \emph{needs\_backport} tags, and these tags are mostly for the \texttt{default} branch rather than version branch pull-requests. Overall, inconsistencies exist on both sides, which can harm both project management and research.

\textbf{Incompatibility.} Some pull-requests need to be changed to make them compatible with \texttt{other} branches when backporting. For example, a backport (i.e., \#24323 \cite{incom_pullrequest24323}) of CPython is changed 2\% (i.e., \code{with os\_helper.temp\_cwd() as cwd:} changed to \newline \code{ with support.temp\_cwd() as cwd}) from pull-request \#23412 \cite{incom_pullrequest23412}.  In our experiment, we used the Python \emph{SequenceMatcher} algorithm (i.e., based on Ratcliff/Obershelp pattern-matching algorithm \cite{ratcliff1988pattern}) and found around 50\% of pull-requests need to be changed with more than 13\% code (i.e., details in Table~\ref{table_challengesbackports}). Changing code can take considerable time and effort to resolve issues while merging \cite{10.1007/s10515-017-0227-0}, which varies for our subject systems. For example, Table~\ref{table_challengesbackports}'s third column shows only 3\% changes required for Kibana's backport where automated approaches are applied for detecting and merging backports. Thus, backports are created and merged within a day (i.e., 1 day to create and merge in third and fourth columns). Less time in creating and merging can be the reason for less incompatibility in Kibana, since versions are less diverted in a short period. Similarly, the Electron repository also shows low incompatibility 04\% with 05 days to create backports and 01 days to merge. The highest incompatibility, 25\%, is shown by the Bootstrap repository, which takes on average 22 days to create and 15 days to merge backports. Likewise, Julia, CPython, show a high level of incompatibility (i.e., 20 and 16, respectively) with 25 to 30 days to create and 7 to 10 days to merge the backports. There can be a correlation between incompatibility and delay. One reason can be developers take time to make the changes compatible while creating and merging backports. If we look into the incompatibility of each change type of backports in Table \ref{table_challengesBackportTypes}, we can see that each type of backports is needed to be changed, and the changes vary from 10\% to 17\%.

\textbf{Creating Delay.} Since creating a backport is an independent task, it might take time. We find that, on average, 16  days are required to create a backport after the original one is merged in Table \ref{table_challengesbackports}. The average creating time varies for different repositories. The fastest repository to create backport is Kibana, and we found in the strategies that automated identification and merging are used in Kibana. On the other hand, Ansible takes 13 days to create backport, and the repository also uses a script to identify backports. Thus, how automated techniques are used in repositories need to be investigated for creating backports in less time. The highest time required to create backports in CPython repository, if we look into the inconsistency column of CPython, we can see that around 97\% backports are not consistent with linking, and their incompatibility rate is also high, around 16\%. Similarly, CMSSW (17 days), Nixpkgs (21 days), Julia (25 days) show a high level of inconsistencies 56.93\%, 60.63\%, 51.38\% with a high level of incompatibilities. Thus, low-level maintenance can be the reason for the delays in creating backports. If we look into Table \ref{table_challengesBackportTypes} of change types, we see creating delays vary for different types of backports. Test changes in backports require a considerable time to create, which is 29 days in Table \ref{table_challengesBackportTypes}. On the other hand, refactor and optimize-backports are created in a short time, 5 and 7 days—more investigation is required to identify the reasons for change-type importance.

\textbf{Merging Delay.} We investigated the time it takes to merge backports. On average, 5 to 6 days are required to merge a backport after its creation (see Table \ref{table_challengesbackports}). ElasticSearch, Electron, and Kibana repositories create and merge backports in a short time (e.g., 1 day to merge). Backport merging time for most repositories is not high, varying between 1 to 7 days. Bootstrap, CPython, and  CMSSW are exceptions, taking 15, 10, and 10 days respectively to merge a backport as they take time to create them with high inconsistency. Furthermore, in Table \ref{table_challengesBackportTypes} we see Document, Refactor and Feature backports are merged in a day, whereas Fix and Test backports require more than five days to merge. Although the average backport merge time (5.40 days) is less than the normal pull-request merge time (8.83 days) it is not negligible given they are already reviewed.

\textbf{Failport.} Similar to a normal pull-request, a backport can also be rejected for many reasons. Our analysis found that at least 10\% of backports are rejected, which is also less than the 20\% rejected normal pull-requests. The lowest number of backports are rejected in Kibana, ElasticSearch repositories as we already recognized they are highly maintained. CPython has the highest percentage of failports with a high percentage of inconsistencies. Thus, maintenance can be the reason for the high number of failures. Change type-wise backports in Table \ref{table_challengesBackportTypes} show more failports for tests and features. Why certain types of backports are failed can be investigated further. Moreover, why reviewed pull-requests are failed should be accounted for investigation by practitioners. Knowing the reasons for failing to port will be worthwhile to resolve some issues.     

\section{Results}
\label{sec_result}

In this section, we summarize our findings on backports while answering our four research questions.

\begin{table}
\centering
\caption{Challenges of change types in backporting}
\resizebox{.85\columnwidth}{!}{%
\begin{threeparttable}
\begin{tabular}{|c|l|l|l|l|l|}
\hline 
 \rot{0}{Type}    & \rot{0}{ INC}      & \rot{0}{IPA}      & \rot{0}{Creating}      & \rot{0}{Merging}  & \rot{0}{ Failport} \\ 
 \rot{0}{}    & \rot{0}{(\%)}      & \rot{0}{(\%)}      & \rot{0}{delay (days)}      & \rot{0}{delay (days)}  & \rot{0}{(\%)} \\ \hline 
Fix              & 51.75          & 15          & 13         & 5      & 07.85  \\
Security     & 52.28          & 13          & 19         & 3      & 05.14  \\
Test             & 53.66          & 17          & 29         & 6      & 10.27  \\
Document         & 56.38          & 10          & 17         & 1      & 06.85  \\
Refactor         & 33.22          & 13          & 05          & 1      & 04.23  \\
Feature          & 64.35          & 10          & 16         & 1      & 09.84  \\
Optimize       & 48.80          & 15          & 07          & 2      & 07.85  \\
\hline          
\end{tabular}
    \begin{tablenotes}
      \small
      \item INC = Inconsistency, IPA = Incompatibility.
      
    \end{tablenotes} 

\end{threeparttable} 
\label{table_challengesBackportTypes}
}
\vspace{-5mm}
\end{table}

\textbf{RQ1.  How are pull-requests backported?} 
To answer RQ1, we summarized the contribution documents in Section \ref{efficiencies}. In general, strategies show, backporting commits are merged by creating new pull-requests. Our data also supports this information; Table~\ref{table_backport_forward_types} shows, on average, 5,203.20 pull-requests are created among the 10 repositories for backporting changes. Maintainers also suggest having backport changes in the \texttt{default} branch first, and we found on average 1,639.20 forwardports in \texttt{default} branches from 10 subject systems. This number could have been higher, but references are missing in many cases (i.e., 51\% of pull-requests were found to be linked). Apart from this, not all changes are propagated with backports, other methods such as cherry-picking and patching can also merge the pull-requests. In general, original pull-requests are forwardported, then backports are created and backported, where references to the original pull-requests are written in backports for future maintenance.

\textbf{RQ2. What kinds of pull-requests are  backported?} 
Table~\ref{table_backport_forward_types} shows the different types of backports.  In contribution documents (i.e., discussed in Section \ref{efficiencies}), we found that fix and security pull-requests are commonly backported. Our experiments on such backports in Table \ref{table_backport_forward_types} show on average 768.10 bug and 103.10 security fixes are backported. We also found that, on average, test (815.10), document (803.60), refactor (30.7), feature (694.90), and optimize (100.60) pull-requests are also backported. As we discussed earlier, pull-requests can be classified in many ways, and many of the types are not common in all repositories. Their nonexistence is represented with zeros in Table~\ref{table_backport_forward_types} if pull-requests do not have specific tags for the change types or the tags are rarely used. The \texttt{Other} column in the table shows a considerable number of backports that represent the pull-requests without any specific change type tags. Our clustering algorithm in Table \ref{table_backporttypes} suggests that they are mostly tagged with versions, releases, components, and community-related labels. In the scope of this study, we only consider tags related to change types, and columns 5 to 11 represent the results in Table \ref{table_backport_forward_types}. Additionally, Table \ref{table_challengesBackportTypes} shows that most backport types have similar challenges. Overall, pull-requests with all types of changes are backported, and bug, test, document, and feature changes backports are more common for most repositories. 

\textbf{RQ3. What strategies are used for backporting?}
Table \ref{table_repotypes} shows tags used in repositories for backporting. Since tags are mostly used to organize issues and pull-requests in repositories, here backports are also organized with tags that include both ``backport" and other tags (i.e., change type tags, version tags). Overall, both backports and forwardports are tagged for \texttt{default} and \texttt{other} branches, and backports are linked with referenced original pull-request numbers. Some repositories have introduced techniques for overcoming backporting challenges, such as automating a part of the process. For example, Ansible and Kibana automate label generation. The outcome is visible in Table \ref{table_challengesbackports}, Kibana repository has a low number of inconsistencies and incompatibilities compared to the other repositories and pull-requests numbers. However, the approaches are logic-based and can be improved. Some studies work on automating commit-merging, but these need to be analyzed for different types of backporting. Since considerable time is required to create and merge backports, automatic identification and merging can reduce delay and effort for both development and management teams. However, current strategies are only concerned with automated labelling, and the benefits of automated merging still need to be investigated.

\textbf{RQ4. What challenges are encountered when backporting?} 
We collected data for each of the six challenges as shown in Section~\ref{sec_practice_challenges}. The inconsistency column represents the data of missing references between backports and original pull-requests in Table \ref{table_challengesbackports}. Around 49\% of backports are not linked to the original pull-requests. Similarly, the incompatibility column represents the code differences of the commit changes from original pull-requests to backports. The incompatibility column shows that merging required considerable change in a backport (on average 13\%). Inconsistencies exist in all repositories from 10.24\% to 96.80\%, and these inconsistencies need to be resolved for future maintenance. On the other hand, incompatibility needs to be resolved just before the merging backports, which varies from 03\% to 25\% among 10 repositories. Apart from these challenges, considerable time is required to create and merge a backport after merging the original pull-request. On average, 15-16 days are required to create a backport and 5-6 days to merge it. If we compare the merging time (i.e., 5.40 days) against normal pull-request merging time (8.83 days), we see it takes less time, and in most of the repositories, backport merging takes less time (i.e., in between 1 to 15 days) than the normal pull-request merging time (i.e., in between 5 to 17 days). But it is not negligible as backports are already reviewed. Besides, the required 16 days to create backports can propagate other issues in a repository. Likewise, the percentage of port changes failure in \texttt{other} branches is only 10.41 on average, and it varies from 03.46\% to 23.20\% among the 10 repositories. This rejection rate is double (20.34) and varies from 12.97 to 46.77 for normal pull-requests. But again, why already reviewed backports are failed to port is worth investigating for future management. Table \ref{table_challengesBackportTypes} also shows similar challenges for the backport types. Moreover, all the change type shows significant inconsistencies, incompatibilities, and delays. So referencing to organize, transforming to make compatible, and waiting to create and merge are the main challenges to overcome for backports.

\vspace{-3mm}
\section{Discussion and Future Direction}
\label{discussion}

This section discusses the implications of our findings for backport practitioners and researchers. Figure \ref{seriesRepo} shows that some projects have many backports while some have a few. Our work here includes projects with at least 400 backports to see strategies and challenges when the backport number is high. However, considering the backport appearance, projects can be further analyzed for the reasons, impacts, and techniques of backports among the high variances. We discussed the frequency of the types; however, discussing their distribution in backports and normal pull-requests can reveal some key characteristics that might be essential for prediction-based approaches. Table \ref{table_challengesbackports} and \ref{table_challengesBackportTypes}'s incompatibility columns show that backports in repositories and change types need to be changed while porting. In our analysis of Table \ref{table_backport_forward_types}, we found that various change types (i.e., not only highly emphasized bug-fix and security changes) are frequently backported. Therefore, what kind of distinct measures are required for each of them can be investigated further. For example, other properties of changes should be investigated rather than just looking for bug or security properties for backporting. Since we are now familiar with the change amount, what changes (e.g., resolving dependencies or others) are desired and how the process can be automated require future exploration. As well, researchers and contributors need to address inconsistencies before analyzing and organizing backports in repositories. The above knowledge can help integrators select appropriate tools for identifying and merging backports.

Backports are already reviewed, but a considerable number of pull-requests fail to be merged. These failports need to be analyzed, and how the acceptance rate is related to project characteristics needs to be explored. Knowledge regarding these can help reviewers to suggest generic solutions for appropriate versions. Similarly, a considerable amount of time is required to create and merge backports. How these can be accomplished in a reasonable time needs to be analyzed with respect to backporting types. Getting this knowledge can help release planners to work on milestones and documentation as relevant artifacts (i.e., pull-requests) need to be identified in release notes \cite{RelContentHuman22} \cite{IJSEKE2021}.
There can be many ways to backport (e.g., patching, cherry-picking, and so on)
and determining their suitability for a project is important for project management decisions. 

We plan to analyze the risk factor of backporting in the open-source community. A more detailed understanding of the characteristics of repositories and pull-requests is needed in order to develop general automatic techniques. Current techniques are logic-based and repository-dependent, whereas more generalized techniques are required for pull-request linking, identifying, and merging.

In support of future directions, we identified backport change types and found that bug, test, document, and feature changes are commonly backported to different versions. As we understand now, if there is a bug or security fix, the probability that the change will be backported is high. Similarly, if new test cases are added, documentation changed, code refactoring performed, optimization introduced, and even a small feature added, backports can be initiated to maintain the quality of the stable versions. On the contrary, large and complex changes are preferred to be discarded for stable branches and need to be investigated for other properties along with the types and challenges. A few backporting challenges included link inconsistencies (49\%), code incompatibilities (13\%), failures (10\%), and delays (e.g., 16 days to create while porting  to \texttt{other} versions). We identified the challenges to indicate obstacles in the backporting process, and they may not present the developer's perspective, and an in-depth survey is required to know the developer's perspective. A few repositories used automated approaches to address some challenges. Our study showed there is evidence for backporting first to the \texttt{default} branch and some correlation among inconsistencies, incompatibilities and delays.

\vspace{-2mm}
\section{THREATS TO VALIDITY}
\label{threatstovalidity}

\textbf{Construct validity.} We collected our data using backport-related keywords and extracted backports if the keywords used in pull-requests occurred more than once. Our assumption for identifying backports with keywords might not be true for a few cases. However, more than 90\% of the pull-requests in the 68,424 backports of our dataset are confirmed to be backports from the label, title tags and analyses. We randomly sampled ten tagged backports to investigate tagging credibility and found that nine of them are ported with only backporting changes for particular tasks. The last one is tagged for porting various changes from a group of pull-requests for a release branch. Thus, our results of the backporting analysis are not subject to any chance. 

\textbf{External validity.} We chose both backports and normal pull-requests from GitHub subject systems. Considering the pull-requests from other environments with respect to their types could have made the results more generalizable. However, we used 68,424 backports and 293,090 normal pull-requests from the 10 projects (i.e., 2 Python, 1 Nix, 1 JavaScript, 1 Java, 1 Julia, 2 C++, 1 PHP, and 1 TypeScript based projects) to generalize our findings. 
Although we have many backports in our dataset, they are collected from mostly large projects. There is a chance that the pull-requests may not portray the pull-requests structure of medium or small projects. However, we believe that the backports are common in large projects, and the results can be generalized to use in such cases from the validated data.

\section{CONCLUSION}
\label{sec_conclusion}

Our exploratory study investigated backports in open-source software, focusing on change types, strategies and challenges. We found there are various pull-request change types (i.e., fix, security, release, test, document, refactor, feature, optimize) that are backported during software maintenance. Major backporting challenges include inconsistent links, incompatible code, and delays. Our future goal is to overcome these challenges by introducing automated techniques for identifying, tracking, and merging backports. In support of this goal and to encourage others, we created a dataset that can be used to propose models for identifying backports. 

\section{DATA AVAILABILITY}
\label{dataA}
Data are released on Zenodo \cite{backportdata}. The provided dataset contains repository, pullID, origin, status, label, discussion, and commit information of normal pull-request and backports in two CSV files. The data can be used for binary classifications (i.e., for identification) and changeset analyses (i.e., for automated merging). 



\section{Acknowledgments}
This research is supported in part by the Natural Sciences and Engineering Research Council of Canada (NSERC) Discovery grants, and by an NSERC Collaborative Research and Training Experience  (CREATE) grant, and by two Canada First Research Excellence Fund (CFREF) grants coordinated by the Global Institute for Food Security (GIFS) and the Global Institute for Water Security (GIWS).



\bibliographystyle{ACM-Reference-Format}
\bibliography{References}


\end{document}